\documentstyle[sprocl]{article}

\def\Journal#1#2#3#4{{#1} {\bf #2}, #3 (#4)}

\def\NIMA{{\em Nucl. Instrum. Methods} A}

\def\be{\begin{equation}}
\def\ee{\end{equation}}
\def\bea{\begin{eqnarray}}
\def\eea{\end{eqnarray}}

\begin{document}

\title{Recent Results from CLEO Collaboration}

\author{Yongsheng Gao}

\address{Physics Department, Southern Methodist University\\
Dallas, TX 75275-0175, USA\\ E-mail: gao@mail.physics.smu.edu} 


\maketitle\abstracts{ 
Recent CLEO results on beauty and charm meson 
decays presented at the International Conference on Flavor Physics
(ICFP2001) are discussed. 
The results include exclusive rare $B$ decays ($B$ $\to$ $PP$, $PV$,
$VV$, $\phi K^{(*)}$, $l^{+}l^{-}K^{(*)}$), inclusive $b \to s \gamma$,
and $CP$ violation measurements in beauty and charm mesons.
Preliminary results on rare $B$ decays using CLEO III data are also
included.
}

\section{Introduction}
The Cornell Electron Storage Ring (CESR) is a symmetric $e^+ e^-$ 
collider operating at the $\Upsilon$(4S) resonance.  
The CLEO II and II.V detector configurations are described in detail 
elsewhere~\cite{cleodetector}.
The integrated luminosity of the full CLEO II+II.V data is 9.2 fb$^{-1}$ 
at the $\Upsilon$(4S) resonance, which corresponds to about 9.7 $\times$ 
10$^6$ $\mbox{$B\bar{B}$}$ pairs, and 4.6 fb$^{-1}$ at energies just below 
the $\mbox{$B\bar{B}$}$ threshold.
The CLEO III data consists of 6.9 fb$^{-1}$ at the $\Upsilon$(4S) resonance 
and 2.3 fb$^{-1}$ at energies just below the $\mbox{$B\bar{B}$}$ threshold.

The origin of $CP$ violation is one of the most important problems
of experimental high energy physics.
The study of $B$ mesons has been attracting extensive world wide
attention because it will allow for a decisive test of the
quark-mixing sector in the Standard Model (SM).
It is very important to test whether the SM provides the 
correct description of $CP$ violation, in order to search for 
new physics beyond the SM.

The analyses results described in this paper are mostly related to 
the $CP$ violation study in beauty and charm meson decays.
All simulated event samples were generated using
GEANT-based simulation of the CLEO detector response.

\section{$B$ $\to$ $PP$, $PV$, $VV$ Decays}
The CLEO collaboration has studied two-body charmless hadronic
decays of $B$ mesons into final states
containing two pseudo-scalar mesons, one pseudo-scalar and one vector meson,
and two vector mesons where the meson can be charged or neutral $\pi$, $K$,
$\rho$ or $K^{*}$ etc.
These charmless hadronic $B$ decays are very important in the study of
$CP$ violation in $B$ meson system, especially in the future measurements
of the $CKM$ angles $\alpha$ and $\gamma$~\cite{babarbook}.

\subsection{$B$ $\to$ $PP$ Results}

The CLEO collaboration has observed two-body charmless hadronic $B$ decays
of $B$ $\to$ $\pi^{+}\pi^{-}$, $K^{\pm}\pi^{\mp}$, $K^{\pm}\pi^{0}$,
$K^{0}\pi^{\pm}$, $K^{0}\pi^{0}$ using CLEO II+II.V data. Upper limits are 
set for other $PP$ decays. The detail of the analyses can be found in 
published CLEO papers~\cite{cleopp,cleopi0pi0}. The results are summarized 
in Tables 1.

\begin{table}[hhh]
\label{table1}
\caption{$B$ $\to$ $PP$ results using full CLEO II+II.V data.}
\begin{center}
\begin{tabular}{|c|c|c|c|c|}
 \hline
    {  Mode}  
 &  {  $N_{\rm sig}$}
 &  {  Sig.}
 &  {  Efficiency} 
 &  {  $BR\times 10^6$}    \\ \hline
    {  $\pi^+\pi^-$}      
 &  {  $20.0^{+7.6}_{-6.5}$}  
 &  {  $4.2\sigma$}      
 &  {  $48\%$} 
 &  {  $4.3^{+1.6}_{-1.4}\pm 0.5$} \\
    {  $\pi^{\pm}\pi^0$} 
 &  {  $21.3^{+9.7}_{-8.5}$}  
 &  {  $3.2\sigma$}      
 &  {  $39\%$}
 &  {  $<12.7\;(90\% {\rm C.L.})$} \\
    {  $\pi^0\pi^0$}     
 &  {  $6.2^{+4.8}_{-3.7}$} 
 &  {  $2.0\sigma$}      
 &  {  $29\%$} 
 &  {  $<5.7\;(90\%{\rm C.L.})$}  \\   \hline
    {  $K^{\pm}\pi^{\mp}$} 
 &  {  $80.2^{+11.8}_{-11.0}$} 
 &  {  $11.7\sigma$}      
 &  {  $48\%$}
 &  {  $17.2^{+2.5}_{-2.4}\pm 1.2$} \\
    {  $K^{\pm}\pi^0$}   
 &  {  $42.1^{+10.9}_{-9.9}$} 
 &  {  $6.1\sigma$}       
 &  {  $38\%$} 
 &  {  $11.6^{+3.0}_{-2.7}{}^{+1.4}_{-1.3}$} \\
    {  $K^0\pi^{\pm}$}   
 &  {  $25.2^{+6.4}_{-5.6}$} 
 &  {  $7.6\sigma$} 
 &  {  $14\%$} 
 &  {  $18.2^{+4.6}_{-4.0}\pm 1.6$} \\
    {  $K^0\pi^0$}       
 &  {  $16.1^{+5.9}_{-5.0}$} 
 &  {  $4.9\sigma$}         
 &  {  $11\%$} 
 &  {  $14.6^{+5.9}_{-5.1}{}^{+2.4}_{-3.3}$} \\ \hline
    {  $K^+K^-$}         
 &  {  $0.7^{+3.4}_{-0.7}$} 
 &  {  $0.0\sigma$}          
 &  {  $48\%$} 
 &  {  $<1.9\;(90\% {\rm C.L.})$} \\
    {  $K^{\pm}K^0$}     
 &  {  $1.4^{+2.4}_{-1.3}$} 
 &  {  $1.1\sigma$}          
 &  {  $14\%$} 
 &  {  $<5.1\;(90\% {\rm C.L.})$} \\
    {  $K^0\bar{K^0}$}   
 &  {  $0$} 
 &  {  $0.0\sigma$}          
 &  {  $5\%$} 
 &  {  $<17\;(90\% {\rm C.L.})$} \\
\hline
\end{tabular}
\end{center}
\end{table}

Recently, preliminary results on these decay modes using CLEO III data
are reported at Lepton Photon 2001 Conference~\cite{cleolp01}. 
The CLEO III data consists of 6.9 fb$^{-1}$ at the $\Upsilon$(4S) resonance 
and 2.3 fb$^{-1}$ at energies just below the $\mbox{$B\bar{B}$}$ threshold.
The preliminary results summarized in Table 2 are based on about half of the
CLEO III data. They are in good agreement with previous published CLEO
results using CLEO II+II.V data (Table 1).

\begin{table}[hhh]
\label{table2}
\caption{Preliminary $B$ $\to$ $PP$ results using part of CLEO III data.}
\begin{center}
\begin{tabular}{|c|c|c|c|c|}
 \hline
    {  Mode}  
 &  {  $N_{\rm sig}$}
 &  {  Sig.}
 &  {  Efficiency} 
 &  {  $BR\times 10^6$}    \\ \hline
    {  $\pi^+\pi^-$}      
 &  {  $3.9^{+1.5}_{-1.2}$}  
 &  {  $2.2\sigma$}      
 &  {  $35\%$} 
 &  {  $3.2^{+3.3+1.0}_{-2.5-1.0}$} \\
    {  $\pi^{\pm}\pi^0$} 
 &  {  $11.5^{+5.6}_{-4.5}$}  
 &  {  $3.4\sigma$}      
 &  {  $29\%$}
 &  {  $11.7^{+5.7+2.2}_{-4.6-2.4}$} \\
    {  $\pi^0\pi^0$}     
 &  {  $2.7^{+2.4}_{-1.6}$} 
 &  {  $2.9\sigma$}      
 &  {  $29\%$} 
 &  {  $<11\;(90\%{\rm C.L.})$}  \\   \hline
    {  $K^{\pm}\pi^{\mp}$} 
 &  {  $29.2^{+7.1}_{-6.4}$} 
 &  {  $5.4\sigma$}      
 &  {  $46\%$}
 &  {  $18.6^{+4.5+3.0}_{-4.1-3.4}$} \\
    {  $K^{\pm}\pi^0$}   
 &  {  $12.9^{+6.5}_{-5.5}$} 
 &  {  $3.8\sigma$}       
 &  {  $32\%$} 
 &  {  $13.1^{+5.8+2.8}_{-4.9-2.9}$} \\
    {  $K^0\pi^{\pm}$}   
 &  {  $14.8^{+4.9}_{-4.1}$} 
 &  {  $6.2\sigma$} 
 &  {  $12\%$} 
 &  {  $35.7^{+12.0+5.4}_{-9.9-6.2}$} \\
    {  $K^0\pi^0$}       
 &  {  $3.0^{+2.9}_{-2.5}$} 
 &  {  $1.6\sigma$}         
 &  {  $8.5\%$} 
 &  {  $10.4^{+10.0+2.9}_{-8.3-2.9}$} \\ \hline
    {  $K^+K^-$}         
 &  {  $1.0^{+2.4}_{-1.7}$} 
 &  {  $0.6\sigma$}          
 &  {  $36\%$} 
 &  {  $<4.5\;(90\% {\rm C.L.})$} \\
    {  $K^{\pm}K^0$}     
 &  {  $0.5^{+1.9}_{-1.1}$} 
 &  {  $0.8\sigma$}          
 &  {  $12\%$} 
 &  {  $<18\;(90\% {\rm C.L.})$} \\
    {  $K^0\bar{K^0}$}   
 &  {  $0.0^{+0.5}_{-0.5}$} 
 &  {  $0.0\sigma$}          
 &  {  $13\%$} 
 &  {  $<13\;(90\% {\rm C.L.})$} \\
\hline
\end{tabular}
\end{center}
\end{table}

\subsection{$B$ $\to$ $PV$ and $VV$ Results}

The CLEO collaboration has observed charmless hadronic $B$ decays
of $B$ $\to$ $\pi^{\pm}\rho^{0}$ and $\pi^{\pm}\rho^{\mp}$ using 
the full CLEO II+II.V data. Upper limits are set for other $PV$ and $VV$ 
decays. The detail of the analyses can be found in 
published CLEO papers~\cite{cleopv,cleovv}. The results are summarized 
in Tables 3.

\begin{table}[hhh]
\label{table3}
\caption{$B$ $\to$ $PV$ and $VV$ results using full CLEO II+II.V data.}
\begin{center}
\begin{tabular}{|c|c|c|}
 \hline
   {  Decay Mode} 
 & {  $BR$ $\times$ 10$^{6}$}  
 & {  Theoretical Prediction  $\times$ 10$^{6}$}  \\ \hline
   {  $\pi^{\pm}\rho^{0}$}    
 & {  $10.4^{+3.3}_{-3.4} \pm 2.1$}      
 & {  0.4 $-$ 13.0}        \\
   {  $\pi^{\pm}\rho^{\mp}$} 
 & {  $27.6^{+8.4}_{-7.4} \pm 4.2$}       
 & {  12 $-$ 93}           \\
   {  $\pi^{0}\rho^{0}$}     
 & {  $<$ 5.5}  
 & {  0.0 $-$ 2.5}         \\ \hline
   {  $K^{\pm}\rho^{0}$}     
 & {  $<$ 17}          
 & {  0.0 $-$ 6.1}         \\
   {  $\pi^{\pm} K^{*0}$}    
 & {  $<$ 16}         
 & {  3.4 $-$ 13.0}        \\
   {  $K^{\pm} K^{*0}$}      
 & {  $<$ 5.3}  
 & {  0.2 $-$ 1.0}         \\ \hline
   {  $\rho^{0}\rho^{0}$}     
 & {  $<$ 4.6 (5.9)}      
 & {  0.54 $-$ 2.5}        \\
   {  $K^{*0}\rho^{0}$}      
 & {  $<$ 13 (19)}      
 & {  0.7 $-$ 6.2}         \\
   {  $K^{*0} \bar{K^{*0}}$} 
 & {  $<$ 8.7 (10)}      
 & {  0.28 $-$ 0.96}        \\ \hline
\end{tabular}
\end{center}
\end{table}

Besides the $B$ $\to$ $VV$ decays in Table 3, CLEO collaboration also searched
and observed $B$ $\to$ $\phi K^{(*)}$ decays which are clear signature of
gluonic penguin process.
The detail of the analyses can be found in~\cite{cleophik}. 
The results are summarized in Table 4.

\begin{table}[hhh]
\label{table4}
\caption{Observation of $B$ $\to$ $\phi K^{(*)}$ decays using full 
         CLEO II+II.V data.}
\begin{center}
\begin{tabular}{|c|c|c|c|c|} \hline
   {  Mode}  
 & {  $N_{\rm sig}$}
 & {  Sig.}
 & {  Efficienty} 
 & {  $BR\times 10^6$}    \\ \hline
   {  $\phi K^{\pm}$}    
 & {  $14.2^{+5.5}_{-4.5}$}  
 & {  $5.4\sigma$}      
 & {  $54\%$} 
 & {  $5.5^{+2.1}_{-1.8}\pm 0.6$} \\
   {  $\phi K^{0}$}     
 & {  $4.2^{+2.9}_{-2.1}$}  
 & {  $2.9\sigma$}      
 & {  $48\%$}
 & {  $<12.3\;(90\% {\rm C.L.})$} \\
   {  $\phi K$ Combined} 
 & {        } 
 & {  $6.1\sigma$}      
 & {        } 
 & {  $5.5^{+1.8}_{-1.5}\pm 0.7$} \\   \hline
   {  $\phi K^{*0}(K^{-}\pi^{+})$}    
 & {  $12.1^{+5.3}_{-4.3}$}  
 & {  $4.5\sigma$}      
 & {  $38\%$} 
 & {  $9.9^{+4.3}_{-3.5}\pm 1.6$} \\
   {  $\phi K^{*0}(K^{0}\pi^{0})$}     
 & {  $5.1^{+3.9}_{-2.8}$}  
 & {  $2.7\sigma$}      
 & {  $20\%$}
 & {  $46.3^{+35.7+5.9}_{-26.0-6.6}$} \\
   {  $\phi K^{*0}$ Combined} 
 & {        } 
 & {  $5.1\sigma$}      
 & {        } 
 & {  $11.5^{+4.5+1.8}_{-3.7-1.7}$}   \\   \hline
   {  $\phi K^{*\pm}(K^{\pm}\pi^{0})$}    
 & {  $3.8^{+4.1}_{-2.8}$}  
 & {  $1.5\sigma$}      
 & {  $25\%$} 
 & {  $9.3^{+10.1+1.7}_{-7.0-1.5}$} \\
   {  $\phi K^{*\pm}(K^{0}\pi^{\pm})$}     
 & {  $4.0^{+3.1}_{-2.2}$}  
 & {  $2.7\sigma$}      
 & {  $32\%$}
 & {  $11.4^{+9.0}_{-6.3}\pm 1.8$} \\
   {  $\phi K^{*\pm}$ Combined} 
 & {       } 
 & {  $3.1\sigma$}      
 & {       } 
 & {  $10.6^{+6.4+1.8}_{-4.9-1.6}$} \\   \hline
   {  $\phi K^{*}$ Combined} 
 & {        } 
 & {  $5.9\sigma$}      
 & {        } 
 & {  $11.2^{+3.6+1.8}_{-3.1-1.7}$} \\  \hline
\end{tabular}
\end{center}
\end{table}

\section{Inclusive $b$ $\to$ $s \gamma$}

Inclusive $b$ $\to$ $s \gamma$ is electroweak-penguin process which
is sensitive to V$_{ts}^{*}$V$_{tb}$ and new physics beyond the SM.
The experimental challenge in this analysis is to suppress huge 
backgrounds from continuum.
Photon candidate in the momentum range of 2.0GeV $<$ E$_{\gamma}$ $<$ 2.7GeV
are selected.
Lepton tag, Event shape variables (using neural net) and ``pseudo 
reconstruction'' techniques are used to suppress the continuum background.
The measurement result, in comparison with prediction in the SM are:

 \begin{itemize}

   \item {${\cal B}$($b$ $\to$ $s\gamma$)
                       = (2.85 $\pm$ 0.35 $\pm$ 0.22) $\times$ 10$^{-4}$ }
   \vskip 0.3cm

   \item {SM prediction: \hskip 0.2cm
                           ${\cal B}$($b$ $\to$ $s\gamma$)
                           = (3.28 $\pm$ 0.33) $\times$ 10$^{-4}$ }
 \end{itemize}

\section{Exclusive $B$ $\to$ $l^{+} l^{-} K$ and $l^{+} l^{-} K^{*}$}

Exclusive Flavor-Changing-Neutral-Current (FCNC) processes:
$B$ $\to$ $l^{+} l^{-} K$ and $l^{+} l^{-} K^{*}$ are highly suppressed
in the SM. 
The branching ratio of these processes in the SM are 
$\sim$ (10$^{-6}$ to 10$^{-7}$)~\cite{smfcnc}.
However, these branching ratios are sensitive to new physics beyond the
SM (SUSY etc)~\cite{nonsmfcnc}.
The important issues in this analysis are:

 \begin{itemize}

   \item { Select Lepton and Kaon (from pion backgrounds) candidates} 

   \vskip 0.1cm

   \item { Suppress Physics Backgrounds:}

    \begin{itemize}
     
      \vskip 0.1cm

      \item { $B$ $\to$ $J/\psi K^{(*)}$
                              where $J/\psi$ $\to$ $e^{+}e^{-}$ or
                              $\mu^{+}\mu^{-}$           }

      \vskip 0.1cm

      \item { $B$ $\to$ $\psi(2S) K^{(*)}$
                              where $\psi(2S)$ $\to$ $e^{+}e^{-}$ or
                              $\mu^{+}\mu^{-}$           }

    \end{itemize}

   \vskip 0.1cm

   \item { Suppress Continuum and other B backgrounds:}

    \begin{itemize}
     
      \vskip 0.1cm

      \item { Event Shape variable, \hskip 0.4cm
                              Missing Energy, etc }

    \end{itemize}

 \end{itemize}

The detail of the analyses can be found in~\cite{cleofcnc}. 
The results are summarized in Table 5.

\begin{table}[hhh]
\label{table5}
\caption{Exclusive Flavor-Changing-Neutral-Current (FCNC)
         $B$ $\to$ $l^{+} l^{-} K$ and $l^{+} l^{-} K^{*}$ results using
         full CLEO II+II.V data.}
 \begin{center}
 \begin{tabular}{|l|c|c|c|} \hline
         {   Decay Mode}      
    &    {   Efficiency}
    &    {   Evts Obsved}           
    &    {   BR UL (90\% CL)}           \\ \hline
         {    $B \to K^{0} e^{+} e^{-}$}     
    &    {    5.3\%}
    &    {    1         }              
    &    {    $<$ 7.6$\times 10^{-6}$ }         \\ \hline
         {    $B \to K^{0} \mu^{+} \mu^{-}$}     
    &    {    4.1\%}
    &    {    0         }              
    &    {    $<$ 7.8$\times 10^{-6}$ }         \\ \hline
         {    $B \to K^{\pm} e^{+} e^{-}$}     
    &    {    16.5\%}
    &    {    1         }              
    &    {    $<$ 2.3$\times 10^{-6}$ }         \\ \hline
         {    $B \to K^{\pm} \mu^{+} \mu^{-}$}     
    &    {    11.1\%}
    &    {    1         }              
    &    {    $<$ 3.4$\times 10^{-6}$ }         \\ \hline
         {     $B \to K  l^{+} l^{-}$}     
    &    {     37.0\%}
    &    {     3         }              
    &    {     $<$ 1.5$\times 10^{-6}$ }         \\ \hline
         {    $B \to K^{*\pm}(K^{0}\pi^{\pm}) e^{+} e^{-}$}     
    &    {    1.9\%}
    &    {    0         }              
    &    {    $<$ 12.8$\times 10^{-6}$ }         \\ \hline
         {    $B \to K^{*\pm}(K^{0}\pi^{\pm}) \mu^{+} \mu^{-}$}
    &    {    1.5\%}
    &    {    0         }              
    &    {    $<$ 15.6$\times 10^{-6}$ }         \\ \hline
         {    $B \to K^{*\pm}(K^{\pm}\pi^{0}) e^{+} e^{-}$}     
    &    {    1.5\%}
    &    {    3         }              
    &    {    $<$ 46.0$\times 10^{-6}$ }         \\ \hline
         {    $B \to K^{*\pm}(K^{\pm}\pi^{0}) \mu^{+} \mu^{-}$}  
    &    {    0.8\%}
    &    {    0         }              
    &    {    $<$ 29.3$\times 10^{-6}$ }         \\ \hline
         {    $B \to K^{*0}(K^{\pm}\pi^{\mp}) e^{+} e^{-}$}  
    &    {    7.1\%}
    &    {    1        }              
    &    {    $<$ 5.0$\times 10^{-6}$ }         \\ \hline
         {    $B \to K^{*0}(K^{\pm}\pi^{\mp}) \mu^{+} \mu^{-}$}  
    &    {    5.2\%}
    &    {    0        }              
    &    {    $<$ 4.6$\times 10^{-6}$ }         \\ \hline
         {    $B \to K^{*0}(K^{0}\pi^{0}) e^{+} e^{-}$}  
    &    {    0.7\%}
    &    {    0        }              
    &    {    $<$ 35.8$\times 10^{-6}$ }         \\ \hline
         {    $B \to K^{*0}(K^{0}\pi^{0}) \mu^{+} \mu^{-}$}  
    &    {    0.2\%}
    &    {    0        }              
    &    {    $<$ 117.3$\times 10^{-6}$ }         \\ \hline
         {     $B \to K^{*}  l^{+} l^{-}$}     
    &    {     18.8\%}
    &    {     4         }              
    &    {     $<$ 2.9$\times 10^{-6}$ }         \\ \hline
\end{tabular}
\end{center}
\end{table}

\section{$CP$ asymmetry measurements in $B$ meson decays}

\subsection{$CP$ asymmetry measurement in inclusive  $b$ $\to$ $s \gamma$}

The $CP$ asymmetry in inclusive  $b$ $\to$ $s \gamma$ defined as: 
    {${\cal A}_{CP} \equiv
     \frac{\Gamma(b\to s\gamma)-\Gamma(\bar{b}\to\bar{s}\gamma)}
     {\Gamma(b\to s\gamma)+\Gamma(\bar{b}\to\bar{s}\gamma)}$}.
is measured at CLEO using full CLEO II+II.V data.
While  ${\cal A}_{CP}$ is predicted in the SM to be $<$ 1.0\%,
${\cal A}_{CP}$ can be as large as $\approx$ (10 $-$ 40)\% due to
non-SM contributions. Therefore, observation of large ${\cal A}_{CP}$
in inclusive  $b$ $\to$ $s \gamma$ would indicate clear signature of
new physics beyond the SM.

Similar to previous inclusive  $b$ $\to$ $s \gamma$ branching ratio
measurement analysis, we select photon candidate in the momentum range of 
2.0GeV $<$ E$_{\gamma}$ $<$ 2.7GeV.
The $s$ quark flavor is tagged by Lepton tag (from the other $B$), or
the flavor of $K$ in ``pseudo reconstruction'' with mistake rates, 
On-off subtraction, particle detection biases taken into account.
The measurement results are:

    \begin{itemize}
     
      \vskip 0.2cm

      \item { ${\cal A}_{CP}$ = ($-$0.079 $\pm$ 0.108 $\pm$ 0.022)(1.0 
              $\pm$ 0.030) \hskip 0.5cm or}

      \vskip 0.2cm

      \item {$-$0.27 $<$ ${\cal A}_{CP}$ $<$ $+$0.10 (90\% C.L.)}

    \end{itemize}

\subsection{$CP$ asymmetry measurement in exclusive rare $B$ decays}

$CP$ asymmetry in self tagging exclusive $B$ decays are also measured
at CLEO using full CLEO II+II.V data~\cite{cleoasy}.
The $CP$ asymmetry is defined as
${\cal A}_{CP} \equiv
            \frac{ {\cal B}(\bar{B} \to \bar{f})-{\cal B}(B\to f) }
                 { {\cal B}(\bar{B} \to \bar{f})+{\cal B}(B\to f)}$.
In SM, the predicted $CP$ asymmetry is $\approx$ $\pm$ 0.1\%~\cite{kramer}.
The results of $CP$ asymmetry measurement in exclusive rare $B$ decays:
$B \to  K^{\pm}\pi^{\mp}$, $K^{\pm}\pi^{0}$, $K^{0}_{s} \pi^{\pm}$,    
$K^{\pm}\eta^{,}$, $\omega \pi^{\pm}$ and charmonia decays:
$B \to J/\psi K^{\pm}$, $\psi(2S) K^{\pm}$ are summarized in Table 6.
The $CP$ asymmetries observed in these decay modes are all consistent
with zero.

\begin{table}[hhh]
\label{table6}
\caption{$CP$ asymmetry measurements for exclusive rare $B$ decays
         using full CLEO II+II.V data.}
\begin{center}
\begin{tabular}{|c|c|c|c|}  \hline
   {  Decay Mode}  
 & {  $N_{\rm sig}$}
 & {  ${\cal A}_{CP}$}
 & {  Prediction}       \\ \hline
   {  $B \to  K^{\pm}\pi^{\mp}$}    
 & {  $80^{+12}_{-11}$}  
 & {  $-$0.04 $\pm$ 0.16}      
 & {  ($+$0.037, + 0.106)}   \\  \hline
   {  $B \to  K^{\pm}\pi^{0}$}    
 & {  $42.1^{+10.9}_{-9.9}$}  
 & {  $-$0.29 $\pm$ 0.23}      
 & {  (+0.026, +0.092) }   \\  \hline
   {  $B \to  K^{0}_{s} \pi^{\pm}$}    
 & {  $25.2^{+6.4}_{-5.6}$}  
 & {  $+$0.18 $\pm$ 0.24}      
 & {  +0.015 }             \\  \hline
   {  $B \to  K^{\pm}\eta^{,}$}    
 & {  $100^{+13}_{-12}$}  
 & {  $+$0.03 $\pm$ 0.12}      
 & {  (+0.020, +0.061) }   \\  \hline
   {  $B \to  \omega \pi^{\pm}$}    
 & {  $28.5^{+8.2}_{-7.3}$}  
 & {  $-$0.34 $\pm$ 0.25}      
 & {  ($-$0.120, +0.024) }  \\  \hline
   {  $B \to  J/\psi K^{\pm}$}    
 & {  $534$}  
 & {  $+$0.018 $\pm$ 0.043}      
 & {  $<$ 0.04 }           \\  \hline
   {  $B \to  \psi(2S) K^{\pm}$}    
 & {  $120$}  
 & {  $+$0.020 $\pm$ 0.092}      
 & {  $<$ 0.04 }           \\  \hline
\end{tabular}
\end{center}
\end{table}

\section{$CP$ asymmetry measurement in charm meson decays}

Cabibbo suppressed charm meson decays have all the necessary ingredients
for {\em CP} violation -- multiple paths to the same final state and a
weak phase difference.  
However, in order to get sizable {\em CP} violation, the 
final state interactions need to contribute non-trivial phase shifts
between the amplitudes, as the SM prediction for $CP$ asymmetry is very 
small ($\sim$ $0.1\%$). 
Therefore, $CP$ asymmetry in charm meson decays is another place
to hunt for new physics beyond the SM.

All of the analyses presented in this paper use the same general technique.
The $D^0$ candidates are reconstructed through the decay sequence
$D^{\star +} \rightarrow D^0 \pi^+_{\rm s}$~\cite{charge}.  
The charge of the slow pion ($\pi^+_{\rm s}$) tags the
flavor of the $D^0$ candidate at production.  The charged daughters
of the $D^0$ are required to leave hits in the silicon vertex detector
and these tracks are constrained to come from a common vertex in three 
dimensions.
The trajectory of the $D^0$ is projected back to its
intersection with the CESR luminous region to obtain the $D^0$
production point.  The $\pi^+_{\rm s}$ is refit with the requirement that
it come from the $D^0$ production point, and the confidence level of the 
$\chi^2$ of this refit is used to reject background.  

The energy release in the $D^\star \rightarrow D^0 \pi^+_{\rm s}$ decay, 
$Q \equiv M^\star - M - m_\pi$, 
obtained from the above technique is observed to have a width of 
$\sigma_Q = 190 \pm 2$ keV,~\cite{qwidth} which is a combination of
the intrinsic width and our resolution,
where $M$ and $M^\star$ are the reconstructed 
masses of the $D^0$ and $D^{\star +}$ candidates respectively, and
$m_\pi$ is the charged pion mass.  The reconstruction technique 
discussed above has also been used by CLEO to measure the $D^{*+}$
intrinsic width, $\Gamma_{D^{*+}} = 96\pm 4\pm 22$ keV 
(preliminary).~\cite{GammaD*}

We searched for direct {\em CP} violation in neutral charm meson decay to 
pairs of light pseudo-scalar mesons: $K^+ K^-$, $\pi^+ \pi^-$, 
$K^0_{\rm s} \pi^0$, $\pi^0 \pi^0$ and $K^0_{\rm s} K^0_{\rm s}$.

\subsection{Search for {\em CP} violation in $D^0 \rightarrow K^+ K^-$
and $D^0 \rightarrow \pi^+ \pi^-$ decay}

The slow charged pion and $D^0$ are produced by the {\em CP}-conserving strong 
decay of the $D^{\star +}$, so the slow pion serves as an unbiased flavor tag
of the $D^0$.  The decay asymmetry can be obtained from the apparent production
asymmetry shown above because the production of $D^{\star \pm}$ is 
{\em CP}-conserving.

The asymmetry result is obtained by fitting the energy release ($Q$) spectrum
of the $D^{\star +} \rightarrow D^0 \pi^+_{\rm s}$ events.  The $D^0$ mass
spectra are fit as a check.  The background-subtracted $Q$ spectrum is fit 
with a signal shape obtained from $K^+ \pi^-$ 
data and a background shape determined 
using Monte Carlo.  
The parameters of the slow pion dominate the $Q$ distribution, 
so all modes have the same shape.
We do the fits in bins of $D^0$ momentum to eliminate any biases due to
differences in the $D^0$ momentum spectra between the data and the MC.
The preliminary results are:

    \begin{itemize}
     
      \vskip 0.2cm

      \item { $A(K^+ K^-) = 0.0005 \pm 0.0218 ({\rm stat}) \pm 0.0084 
              ({\rm syst})$}

      \vskip 0.2cm

      \item {$A(\pi^+ \pi^-) = 0.0195 \pm 0.0322 ({\rm stat}) \pm 0.0084 
             ({\rm syst})$}

    \end{itemize}
 
The measured asymmetries are consistent with zero, and no {\em CP} violation
is seen.  These results are the most precise to date.~\cite{OLDCP}

\subsection{Search for {\em CP} Violation in $D^0 \rightarrow K^0_{\rm
S} \pi^0$, $D^0 \rightarrow \pi^0 \pi^0$ and $D^0 \rightarrow K^0_{\rm s} 
K^0_{\rm s}$ decay}

This analysis~\cite{jaffe} differs from the other analyses 
presented in this paper in 
some of its reconstruction techniques and in the data set used.  
The $\pi^0 \pi^0$ and
$K^0_{\rm s} \pi^0$ final states do not provide sufficiently precise
directional information about their parent $D^0$ to use the intersection 
of the $D^0$ 
projection and the CESR luminous region to refit the
slow pion as described in the general experimental technique section. 
The $K^0_{\rm s} K^0_{\rm s}$ final state is treated the same for consistency.
This analysis uses the data from both the CLEO II and CLEO II.V
configurations of the detector.

The $K^0_{\rm s}$ and $\pi^0$ candidates are constructed using only good 
quality tracks and showers.  The tracks (showers) whose combined 
invariant mass is close to the $K^0_{\rm s}$ ($\pi^0$) mass are
kinematically constrained to the $K^0_{\rm s}$ ($\pi^0$) mass, improving the 
$D^0$ mass resolution.  The tracks used to form $K^0_{\rm s}$
candidates are required to satisfy criteria designed to reduce
background from $D^0 \rightarrow \pi^+ \pi^- X$ decays and combinatorics.  
Candidate events with
reconstructed $D^0$ masses close to the known $D^0$ mass are selected to
determine the asymmetry. 
The total number of $D^0$ and $\overline{D^0}$ candidates for a given final
state is determined as follows.  We fit the $Q$ distribution outside of 
the signal region and interpolate the fit under the signal peak to 
determine the background in the signal region.  We subtract the background
in the signal region from the total number of events there to determine
the total number of signal events.
After background subtraction, we obtain $9099 \pm 153$ $K^0_{\rm s}
\pi^0$ candidates, $810 \pm 89$ $\pi^0 \pi^0$ candidates, and $65 \pm 14$ 
$K^0_{\rm s} K^0_{\rm s}$ candidates.  

The difference in the number of $D^0$ and $\overline{D^0}$ to a given final 
state is determined by taking the difference of the number of events in
the signal region, and the asymmetry is obtained by dividing by the
number of candidates determined above.  This method of determining the
asymmetry implicitly assumes that the background is symmetric.  

We obtain the results:

    \begin{itemize}
     
      \vskip 0.2cm

      \item { $A(K^0_{\rm s} \pi^0) = (+0.1 \pm 1.3)\%$}

      \vskip 0.2cm

      \item {$A(\pi^0 \pi^0) = (+0.1 \pm 4.8)\%$}

      \vskip 0.2cm

      \item {$A(K^0_{\rm s} K^0_{\rm s}) = (-23 \pm 19)\%$ }

    \end{itemize}
 where the uncertainties contain the 
combined statistical and systematic uncertainties.
All measured asymmetries are consistent with zero and no indication of
significant {\em CP} violation is observed.  This measurement of 
$A(K^0_{\rm s} \pi^0)$ is a significant improvement over previous results, 
and the other two asymmetries reported are first measurements.


\begin{thebibliography}{99}
\bibitem{charge} Charge conjugation is implied throughout, except where the 
        charge conjugate states are explicitly shown, such as in an asymmetry
        definition.
\bibitem{qwidth} This result is for the $D^0 \rightarrow K^+ \pi^-$ mode.  
        Other modes have similar widths since the uncertainty on the slow pion 
        dominates the width of the $Q$ distribution.
\end{thebibliography}
\end{document}